\documentclass[12pt]{article}
\usepackage[polish,english]{babel}
\usepackage[latin1]{inputenc}
\usepackage{graphicx}
\usepackage{times}
\usepackage{bbold}
\usepackage[T1]{fontenc}
\usepackage{epsfig}

\def\rf#1{(\ref{eq:#1})}
\def\lab#1{\label{eq:#1}}

\def\br{\begin{eqnarray}}
\def\er{\end{eqnarray}}
\def\be{\begin{equation}}
\def\ee{\end{equation}}
\def\({\left(}
\def\){\right)}

\def\rlx{\relax\leavevmode}
\def\IR{\rlx\hbox{\rm I\kern-.18em R}}

\def\vt{\vartheta}
\def\u2{\mid u\mid^2}
\newcommand{\sbr}[2]{\left\lbrack\,{#1}\, ,\,{#2}\,\right\rbrack}

\begin{document}

\title{Numerical vortex solutions in (3+1) dimensions for the extended $CP^N$ Skyrme-Faddeev model}

\maketitle

\begin{center}
  Pawe\l~Klimas$^{a,}$\footnote{e-mail: {\tt klimas.ftg@gmail.com}} and
  Nobuyuki Sawado$^{b,}$\footnote{e-mail: {\tt sawado@ph.noda.tus.ac.jp}}

\vspace{.5 in}
\small

\par \vskip .2in \noindent
$^{a}$ Instituto de F\'\i sica de S\~ao Carlos; IFSC/USP;\\
Universidade de S\~ao Paulo - USP \\ 
Caixa Postal 369, CEP 13560-970, S\~ao Carlos-SP, Brazil\\
$^b$ Department of Physics, Tokyo University of Science,\\
 Noda, Chiba 278-8510, Japan

\end{center}

\begin{abstract}
We construct numerical vortex solutions in a (3+1) dimensional Min\-kow\-ski space-time
for the extended version of the Skyrme-Faddeev model
with target space $CP^N$. The solutions are essentially composed of $N$-th single 
vortex which does not belong to the integrable sector. They have finite energy per unit length and contain waves pro\-pa\-ga\-ting along vortices with 
the speed of light. 
In this Letter we report on the case $N=2$ and show existence of the solutions with the charges $(n_1,n_2)=(1,2)$.
\end{abstract}

\section{Introduction}

The Skyrme-Faddeev model is an example of a field theory that supports the finite-energy knotted solitons. The significance of this model has increased noticeably when it has been conjectured that the model can be seen as a low-energy effective classical model of the underlying Yang-Mills theory \cite{sf}.  Similarly to many other models \cite{coleman} the classical soliton solutions of the Skyrme-Faddeev model can play a role of adequate normal models useful in description of the strong coupling sector of the Yang-Mills theory. The exact soliton (vortex) solution of the model has been found within the integrable sector \cite{vortexlaf}. Such a sector exists in the version of the model that is an extension of the standard Skyrme-Faddeev model obtained by including some quartic term different to the Skyrme term. The study of the extended models have been originally motivated by the results of the analysis of the Wilsonian action of the $SU(2)$ Yang-Mills theory \cite{gies}. It has been shown that also in the case of the complex projective target space $CP^N$ the extended Skyrme-Faddeev model 
 (in which it has been imposed some special constraints for the parameters of the model) 
possesses an exact soliton solutions in the integrable sector \cite{fk,fkz}. The aim of this Letter is to investigate if the model possesses also some solutions 
outside the integrable sector, i.e., without the constraints. 
The study of such models is promising and could be important for understanding some aspects of the strong coupling sector of the Yang-Mills theory.

\section{The formulation of the model}

The Skyrme-Faddeev model and its extensions on the $CP^1$ target space are usually expressed in terms of the real unit vector $\vec n$. The dimension of a target space is simply related to the number of degrees of freedom of the model. For instance, the model with the $CP^1$ target space has only two independent degrees of freedom. In order to add more degrees of freedom one can consider some higher dimensional target spaces.  The target space (coset space) in the case of some higher dimensional $SU(N)$ Lie groups, i.e. $N>2$, can be chosen in several nonequivalent ways. 

Recently it has been proposed some formulation of the extended Skyrme-Faddeev model on the $CP^N$ target space \cite{fk}. The coset space $CP^N=SU(N+1)/SU(N)\otimes U(1)$ is an example of a symmetric space and it can be naturally parameterized in terms of so called {\it principal variable} $X(g)=g\sigma(g)^{-1}$, with $g\in SU(N+1)$ and $\sigma$ being the order two automorphism under which the subgroup $SU(N)\otimes U(1)$ is invariant i.e. $\sigma(h)=h$ for $h\in SU(N)\otimes U(1)$. The coordinate $X(g)$ defined above satisfies $X(gh)=X(g)$.
 
We shall consider the field theory in $(3+1)$ dimensions defined by the Lagrangian
\br
{\cal L}&=& -\frac{M^2}{2}\,{\rm Tr}\(X^{-1}\,\partial_{\mu}X\)^2
+\frac{1}{e^2}\,{\rm
  Tr}\(\sbr{X^{-1}\,\partial_{\mu}X}{X^{-1}\,\partial_{\nu}X}\)^2 \nonumber\\&+&
\frac{\beta}{2}\, \left[{\rm Tr}\(X^{-1}\,\partial_{\mu}X\)^2\right]^2
+\gamma\, \left[{\rm Tr}\(X^{-1}\,\partial_{\mu}X\;\;
X^{-1}\,\partial_{\nu}X\)\right]^2-\mu^2V
\label{actionx}\nonumber\\
\er
where $M$ is a coupling constant with dimension of mass whereas the coupling constants $e^{-2}$, $\beta$, $\gamma$ are dimensionless. The first term is quadratic in $X$ and corresponds with the Lagrangian of the $CP^N$ model. The quartic term proportional to $e^{-2}$ is the Skyrme term whereas other quartic terms constitute the extension of the standard Skyrme-Faddeev model. The novelty of the model (\ref{actionx}) comparing with that introduced in \cite{fk} is the presence of the potential $V$. The $CP^1$ Skyrme-Faddeev model with the potential have been studied by many authors. The recent results shown that the extended $CP^1$ Skyrme-Faddeev model in (3+1) dimensions possesses some non-holomorphic solutions that do not belong to the integrable sector \cite{fjst}. Since the extended Skyrme-Faddeev model on the $CP^N$ target space also possesses the integrable sector as well as the exact vortex solutions the natural question is if there exist any solutions that do not belong to the integrable sector? In similarity to the paper \cite{fjst} we study such a possibility in the presence of the potential. As it has been explained below  such solutions can be obtained numerically for some choice of the potential. 

\subsection{The parametrization}
Let us shortly discuss the parameterization of the model. According to the previous paper \cite{fk} one can parametrize the model in terms of $N$ complex fields $u_i$, where $i=1,\ldots,N$. Assuming $(N+1)$-dimensional defining representation where the $SU(N+1)$ valued element $g$ is of the form
\br
g\equiv \frac{1}{\vt}\,\(\begin{array}{cc}
\Delta&i\,u\\
i\,u^{\dagger}&1
\end{array}\) \qquad\qquad \qquad \quad \vt\equiv\sqrt{1+u^{\dagger}\cdot u}
\lab{gdef}
\er
and where $\Delta$ is the hermitian $N\times N$-matrix
\br
\Delta_{ij}=\vt\,\delta_{ij}-\frac{u_i\,u_j^*}{1+\vt}
\quad\quad 
\mbox{\rm which satisfies} \quad \quad 
\Delta\cdot u= u\; ; \qquad u^{\dagger}\cdot \Delta= u^{\dagger}
\label{deltadef}
\er
one can express $X(g)=g\sigma(g)^{-1}=g^2$ in terms of $u_i$. The Lagrangian (\ref{actionx}) reads 
\be
{\cal L}=
-\frac{1}{2}\,\left[M^2\,\eta_{\mu\nu}+C_{\mu\nu}\right]\,\tau^{\nu\mu}-\mu^2V
\lab{actioncmunu}
\ee
where the symbols $C_{\mu\nu}$ and $\tau_{\mu\nu}$ read 
\be
C_{\mu\nu}\equiv M^2\,\eta_{\mu\nu}
-\frac{4}{e^2}\left[\(\beta\,e^2-1\)\,\tau_{\rho}^{\rho}\,\eta_{\mu\nu}
+\(\gamma\,e^2-1\)\,\tau_{\mu\nu}+\(\gamma\,e^2+2\)\,\tau_{\nu\mu}\right],
\lab{cmunudef}
\ee
\be
\tau_{\mu\nu}\equiv-\frac{4}{\vartheta^4}\left[\vartheta^2\partial_{\nu}u^{\dagger}\cdot \partial_{\mu}u-(\partial_{\nu}u^{\dagger}\cdot u)(u^{\dagger}\cdot\partial_{\mu}u) \right].
\ee
We shall discuss the specific form of the potential in the further part of the paper. It is enough to assume that $V=V(u^{\dagger},u)$. The variation with respect to $u_i^*$ leads to the equations 
\begin{eqnarray}
&&(1+u^{\dagger}\cdot u)\partial^{\mu}(C_{\mu\nu}\partial^{\nu}u_i)-C_{\mu\nu}\left[(u^{\dagger}\cdot\partial^{\mu}u)\partial^{\nu}u_i+(u^{\dagger}\cdot\partial^{\nu}u)\partial^{\mu}u_i\right]+\nonumber\\
&&+\mu^2\frac{u_i}{4}(1+u^{\dagger}\cdot u)^2\left[\frac{\delta V}{\delta |u_i|^2}+\sum_{k=1}^N|u_k|^2\frac{\delta V}{\delta |u_k|^2}\right]=0.
\label{eom1}
\end{eqnarray}
In order to get the result (\ref{eom1}) one needs to multiply the equation obtained directly from the variation with respect to 
$u^*_i$ by the inverse matrix of $\Delta^2_{ki}$ i.e. $\Delta^{-2}_{ij}=\frac{1}{1+u^{\dagger}\cdot u}(\delta_{ij}+u_i u^{\dagger}_j)$. It leads to the term being a combination of partial derivatives of the potential. We introduce the dimensionless coordinates $(t,\rho,\varphi,z)$ defined as
\br
x^0=r_0t,\quad x^1=r_0\rho\cos\varphi,\quad x^2=r_0\rho\sin\varphi\quad x^3=r_0 z
\er
where the length scale $r_0$ is defined in terms of coupling constants $M^2$ and $e^2$ i.e.
$$
r_0^2=-\frac{4}{M^2e^2}
$$
and the light speed is $c=1$ in the chosen unit system. The linear element $ds^2$ reads
$$
ds^2=r_0^2(dt^2-dz^2-d\rho^2-\rho^2d\varphi^2).
$$
The family of exact vortex solutions has been found for the model without potential $\mu^2=0$ where in addition the coupling constants satisfy the condition $\beta e^2+\gamma e^2=2$. The exact solutions have the form of vortices which depend on some specific combination of the coordinates i.e. one light-cone coordinate $x^3+x^0$ and one complex coordinate $x^1+ix^2$. The functions $u_i(x^3+x^0,x^1+ix^2)$ satisfy the zero curvature condition $\partial_{\mu}u_i\partial^{\mu}u_j=0$ for all $i,j=1,\ldots,N$ and therefore one can construct the infinite set of conserved currents. The interesting problem is the existence of solutions outside the integrable sector and without restriction on the coupling constants. This is the main point of the present paper. We are looking for the solutions that have still dependence on one of the light-cone coordinates but which are not restricted to the holomorphic sector. It has been shown in \cite{fjst} that in the case of $CP^1$ target space and the presence of the potential such solutions exist. In the present case of the $CP^N$ target space we shall consider the following ansatz
\br
u_j=f_j(\rho)e^{i(n_j\varphi+k_j\psi(y))}
\er
where $\psi(y)$ is a real function of the light-cone coordinate and $f_i(\rho)$ is $N$-th element set of real functions. The constants $n_i$ form the set of integer numbers and $k_i$ are some real constants. We shall use the matrix notation for the convenience therefore we define two diagonal matrices
\br
\lambda\equiv{\rm diag}(n_1,\ldots,n_N),\qquad\sigma\equiv{\rm diag}(k_1,\ldots,k_N).
\er 
In the matrix form the ansatz reads $u=f(\rho)\exp{[i(\lambda\varphi+\sigma\psi(y))]}$ where $y$ is either $z+t$ or $z-t$. 
The components of $\tau_{\mu\nu}$ have the following form
\br
\begin{array}{ll}
\tau_{\rho\rho}=\theta(\rho)&\tau_{tt}=(\partial_t\psi)^2\chi(\rho)\\
\tau_{\varphi\varphi}=\omega(\rho)&\tau_{zz}=(\partial_z\psi)^2\chi(\rho)\\
\tau_{\varphi\rho}=-\tau_{\rho\varphi}=i\zeta(\rho)&\tau_{tz}=\tau_{zt}=(\partial_t\psi)(\partial_z\psi)\chi(\rho)\\
\tau_{t\rho}=-\tau_{\rho t}=(i\partial_t\psi)\xi(\rho)&\tau_{z\rho}=-\tau_{\rho z}=(i\partial_z\psi)\xi(\rho)\\
\tau_{t\varphi}=\tau_{\varphi t}=(\partial_t\psi)\eta(\rho)&\tau_{z\varphi}=\tau_{\varphi z}=(\partial_z\psi)\eta(\rho)
\end{array} 
\er
where we have defined the following functions 
\br
\theta(\rho)&:=&-\frac{4}{\vt^4}\,\left[\vt^2\,f'^T.f'-(f'^T.f)(f^T.f')\right]\nonumber\\
\omega(\rho)&:=&-\frac{4}{\vt^4}\,\left[\vt^2\,f^T.\lambda^2.f-(f^T.\lambda.f)^2\right]\nonumber\\
\zeta(\rho)&:=&-\frac{4}{\vt^4}\,\left[\vt^2\,f'^T.\lambda.f-(f^T.\lambda.f)(f'^T.f)\right] \nonumber\\
\xi(\rho)&:=&-\frac{4}{\vt^4}\,\left[\vt^2\,f'^T.\sigma.f-(f^T.\sigma.f)(f'^T.f)\right] \nonumber\\
\eta(\rho)&:=&-\frac{4}{\vt^4}\,\left[\vt^2\,f^T.\lambda.\sigma.f-(f^T.\lambda.f)(f^T.\sigma.f)\right]\nonumber\\
\chi(\rho)&:=&-\frac{4}{\vt^4}\,\left[\vt^2\,f^T.\sigma^2.f-(f^T.\sigma.f)(f^T.\sigma.f)\right].\nonumber
\er
where the derivative with respect to $\rho$ has been denoted by $\frac{d}{d\rho}='$. The equations of motion written in dimensionless coordinates take the form
\br
&&(1+f^T.f)\left[\frac{1}{\rho}\left(\rho\,\tilde{C}_{\rho\rho}f'_k\right)'+\frac{i}{\rho}\left(\frac{\tilde{C}_{\rho\varphi}}{\rho}\right)'(\lambda.f)_k-\frac{1}{\rho^4}\tilde{C}_{\varphi\varphi}(\lambda^2.f)_k\right]\nonumber\\
&&-2\left[\tilde{C}_{\rho\rho}(f^T.f')f'_k-\frac{1}{\rho^4}\tilde{C}_{\varphi\varphi}(f^T.\lambda.f)(\lambda.f)_k\right]\nonumber\\
&&+\frac{\mu^2r_0^2}{M^2}\frac{f_k}{4}(1+f^{T}.f)^2\left[\frac{\delta V}{\delta f_k^2}+\sum_{i=1}^Nf_i^2\frac{\delta V}{\delta f_i^2}\right]=0\nonumber\\
\er
for each $k=1,\ldots,N$, where we have introduced the symbols $\tilde {C}_{\mu\nu}\equiv C_{\mu\nu}/r_0^2M^2$. These of components $\tilde{C}_{\mu\nu}$ which appear in the equations of motion read
\br
&&\tilde{C}_{\rho\rho}=-1+(\beta e^2-1)\left(\theta+\frac{\omega}{\rho^2}\right)+(2\gamma e^2+1)\theta\nonumber\\
&&\tilde{C}_{\varphi\varphi}=-\rho^2+\rho^2(\beta e^2-1)\left(\theta+\frac{\omega}{\rho^2}\right)+(2\gamma e^2+1)\omega\nonumber\\
&&\tilde{C}_{\varphi\rho}=-\tilde{C}_{\rho\varphi}=-3i\zeta\,.
\er

\begin{figure}[t]
\includegraphics[width=8cm,clip]{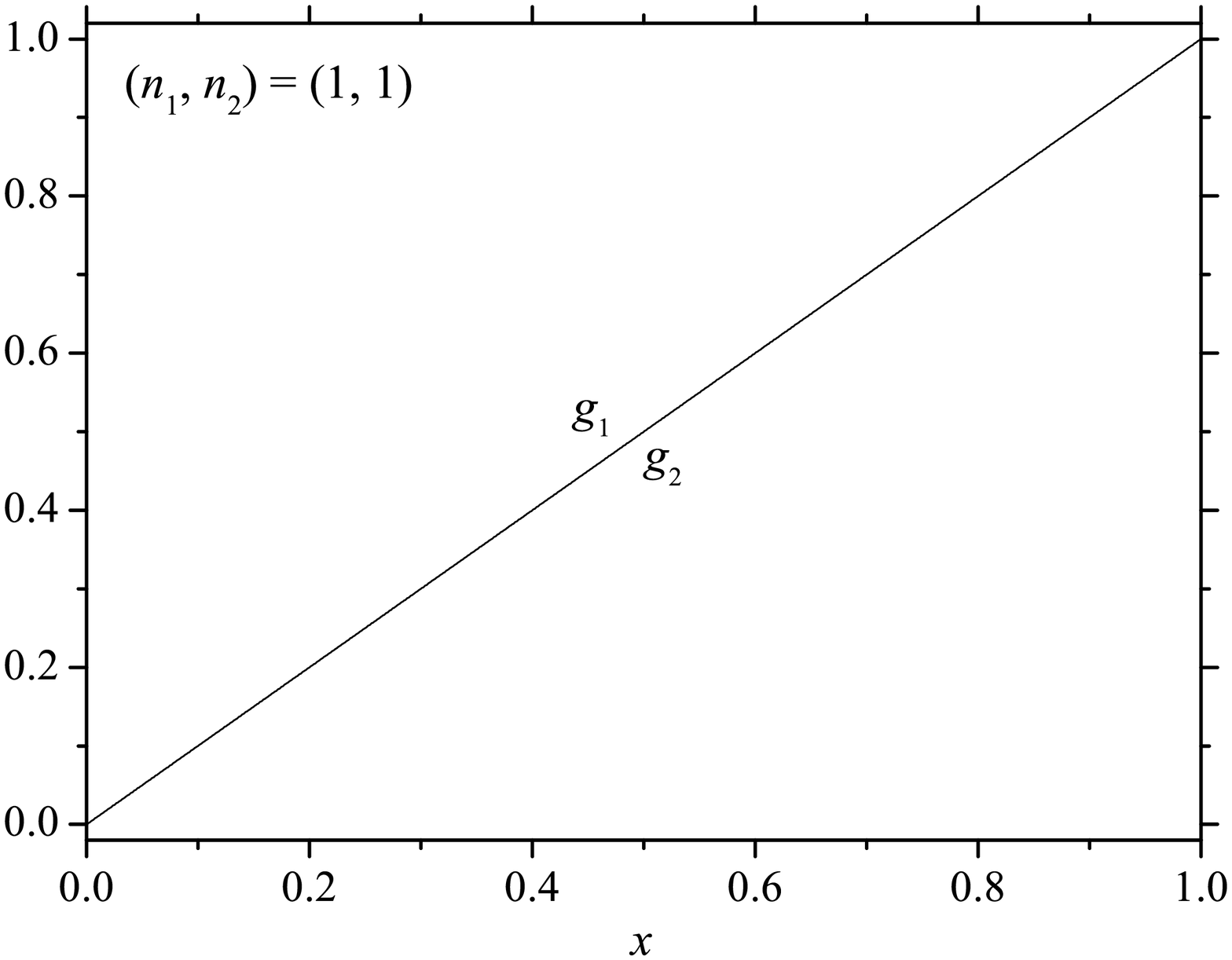}\hspace{-1cm}
\includegraphics[width=8cm,clip]{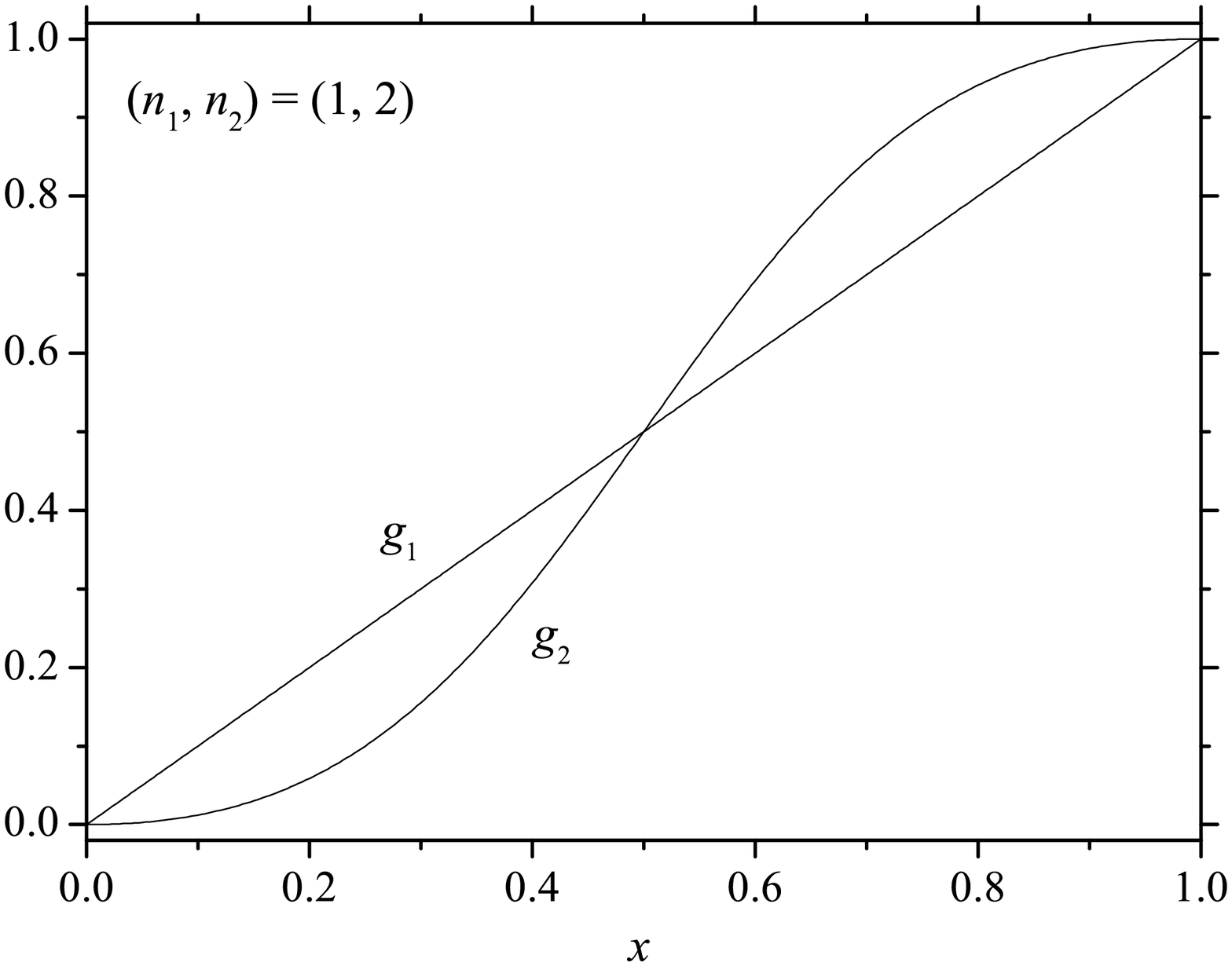}

\caption{\label{profile0}Plot of the profile functions of the solutions from the integrable sector for $(n_1,n_2)=(1,1)$
(left) and $(n_1,n_2)=(1,2)$ (right) . 
The corresponding parameters are  $\beta e^2=1.0, \gamma e^2=1.0$, $\tilde{\mu}^2=0.0$ and an initial 
condition $\alpha_i=1$.}
\end{figure}

\begin{figure}[t]
\includegraphics[width=12cm,clip]{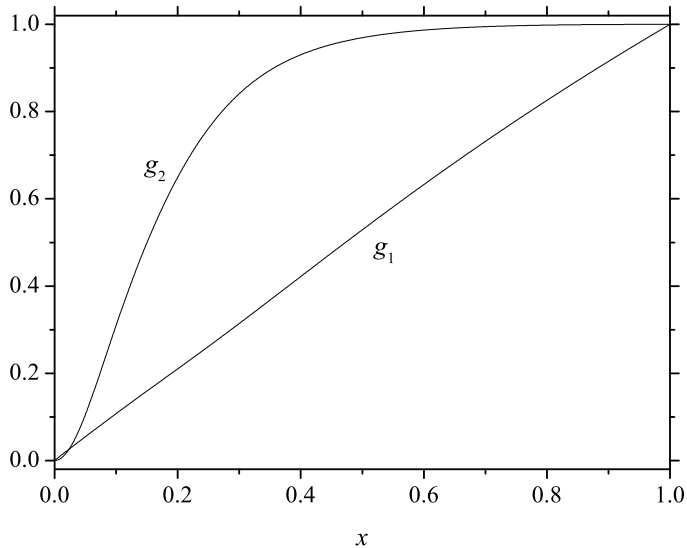}

\caption{\label{profile1}Plot of the profile functions of the solutions not belonging to the 
integrable sector for $(n_1,n_2)=(1,2)$ . The parameters are chosen $\beta e^2=2.0, \gamma e^2=2.0$, 
$\tilde{\mu}^2=1.0$.}
\end{figure}

\subsection{The potential}

There is a plenty of possibilities for the choice of the potential. 
In the case of the Skyrme-Faddeev model with the $CP^1$ target space, usually 
the basic ingredient is a third component of the isovector $\vec n$.  In the present case we have the principal variable $X$ so one can consider the expressions containing trace of $X$ or some other traces involving $X$. In fact the complex fields $u_i$ can be expressed in terms of such expressions.  We shall consider here the potential already expressed in terms of $u_i$.
The possible candidate for the case of the $CP^N$ target space is to introduce a quantity which has a similar role as $n_3$ in the $CP^1$ case. One can consider for instance the expression
\begin{eqnarray}
\mathcal{N}:=\frac{1}{N}\sum_{i=1}^{N}\frac{1-|u_i|^2}{1+|u_i|^2}
=\frac{1}{2}\biggl(\frac{1-|u_1|^2}{1+|u_1|^2}+\frac{1-|u_2|^2}{1+|u_2|^2}\biggr)
\lab{nfac}
\end{eqnarray}
which is suitable for construction of the potential in the $CP^N$ (here in particular $CP^2$) case. It exhibits the following boundary behavior
\br
&&r\to 0~~~~|u_i|\to0~~\Longrightarrow~~\mathcal{N}=1 \nonumber \\
&&r\to\infty~~~~|u_i|\to\infty~~\Longrightarrow~~\mathcal{N}=-1\,.
\er
The expression $\mathcal{N}$ has been chosen in the way that the potential
\br
V&:=&(1-\mathcal{N})^a(1+\mathcal{N})^b \nonumber \\
&=&\frac{(|u_1|^2+|u_2|^2+2|u_1|^2|u_2|^2)^a(2+|u_1|^2+|u_2|^2)^b}{(1+|u_1|^2)^{a+b}(1+|u_2|^2)^{a+b}}
\lab{potential}
\er
with $a\ge 0$, $b>0$,
has some finite (but nonzero) value in the case when one of fields $u_i$ vanishes or grows infinitely. Such behavior is desired from the point of view of numerical stability of the solutions.  
It turns out that such a choice is sufficient to show the existence of some solutions which do not belong to the integrable sector. The analysis of many other possible forms of the potential is out of the scope of the present Letter.

\begin{figure}[t]
~~\\
\hspace{-1cm}\includegraphics[width=8cm,clip]{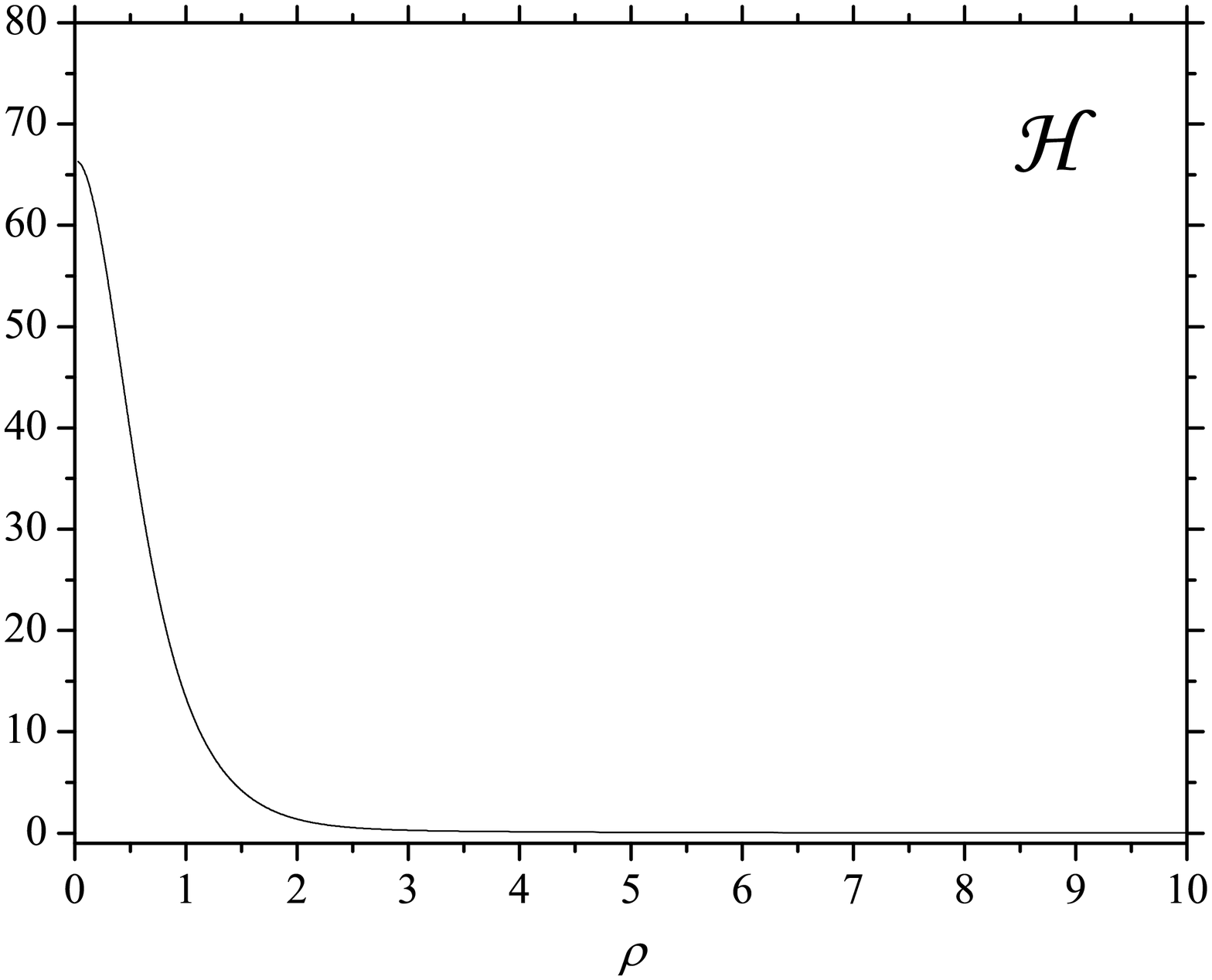}
\includegraphics[width=8cm,clip]{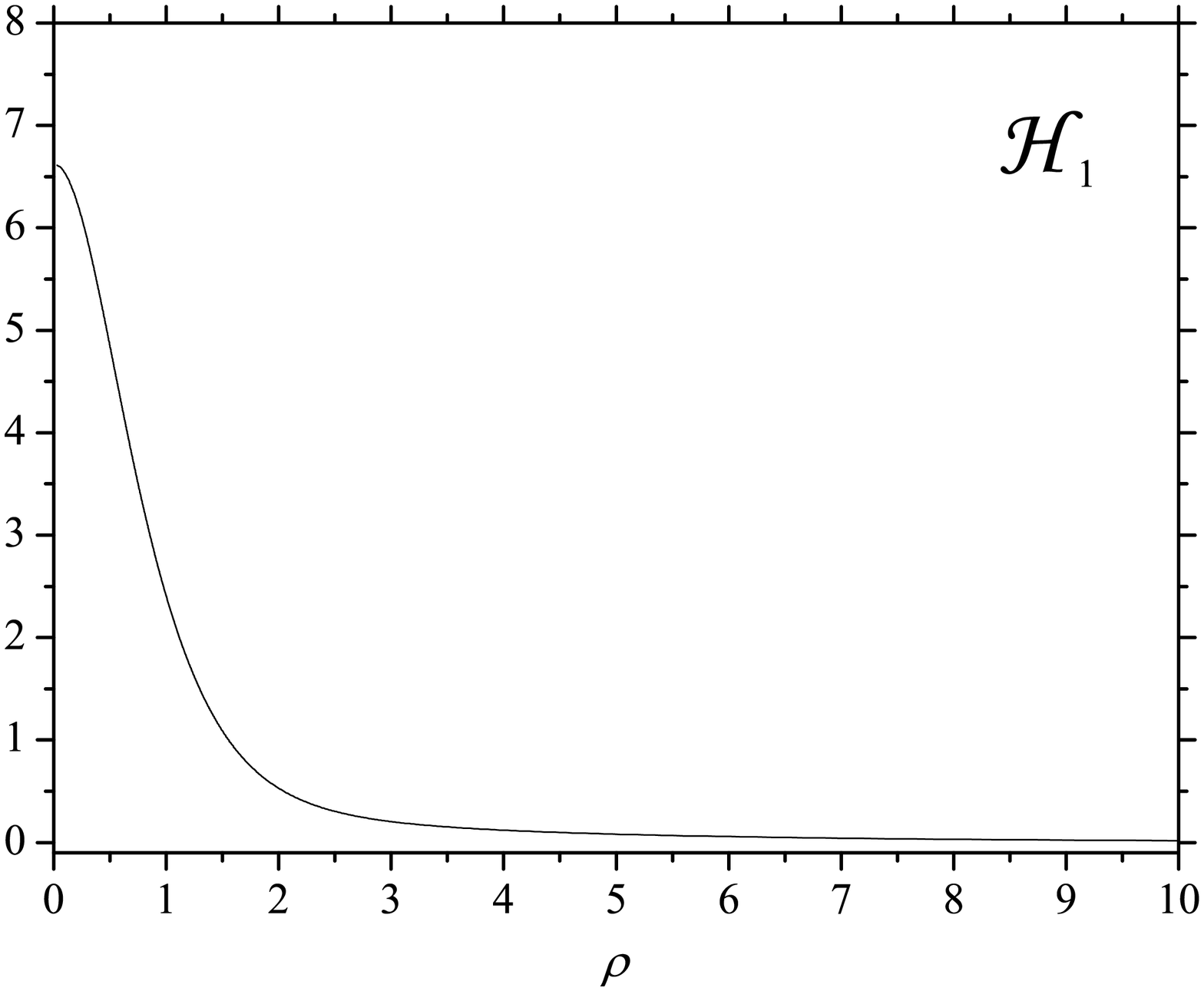}\\
\hspace{-1cm}\includegraphics[width=8cm,clip]{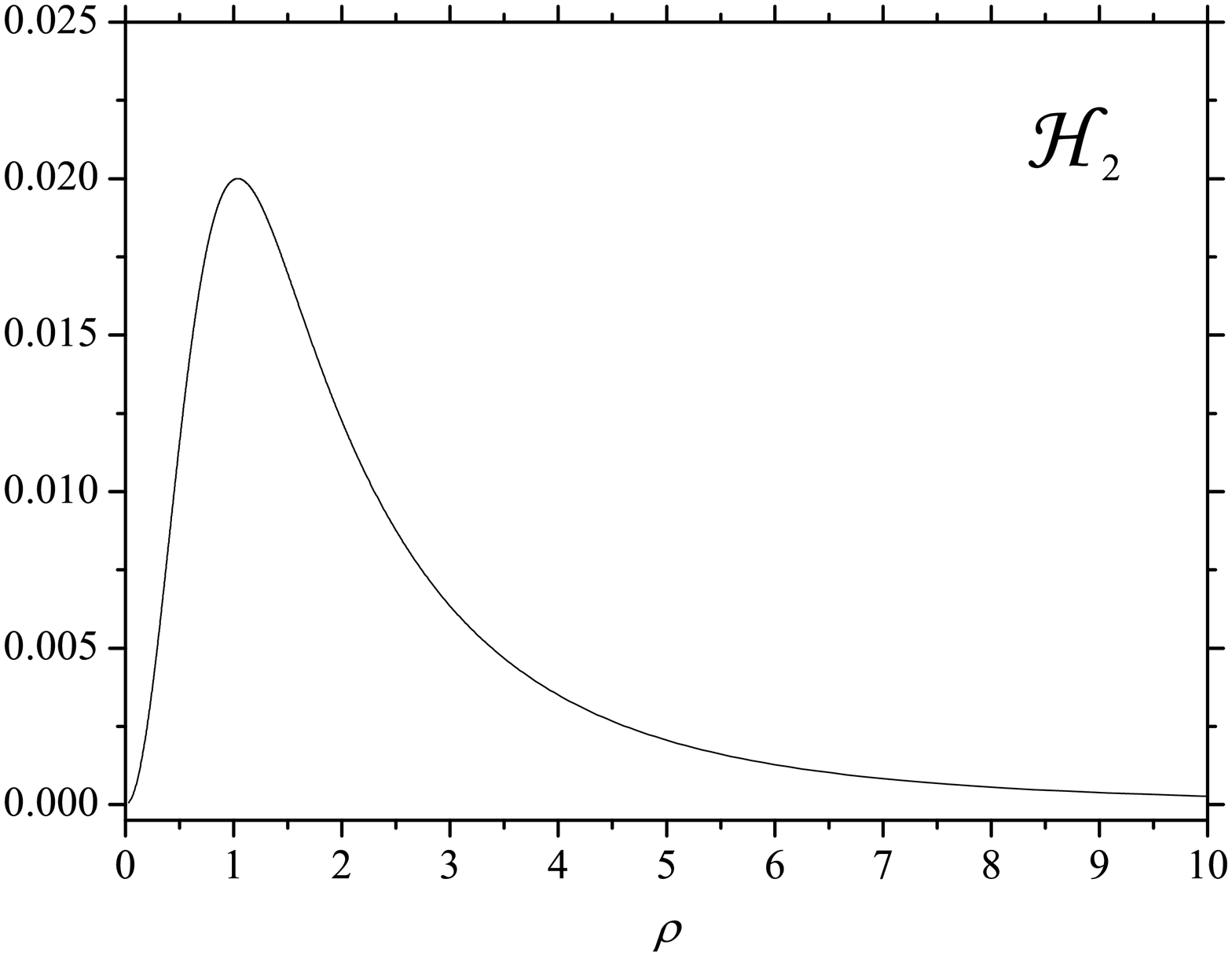}
\includegraphics[width=8cm,clip]{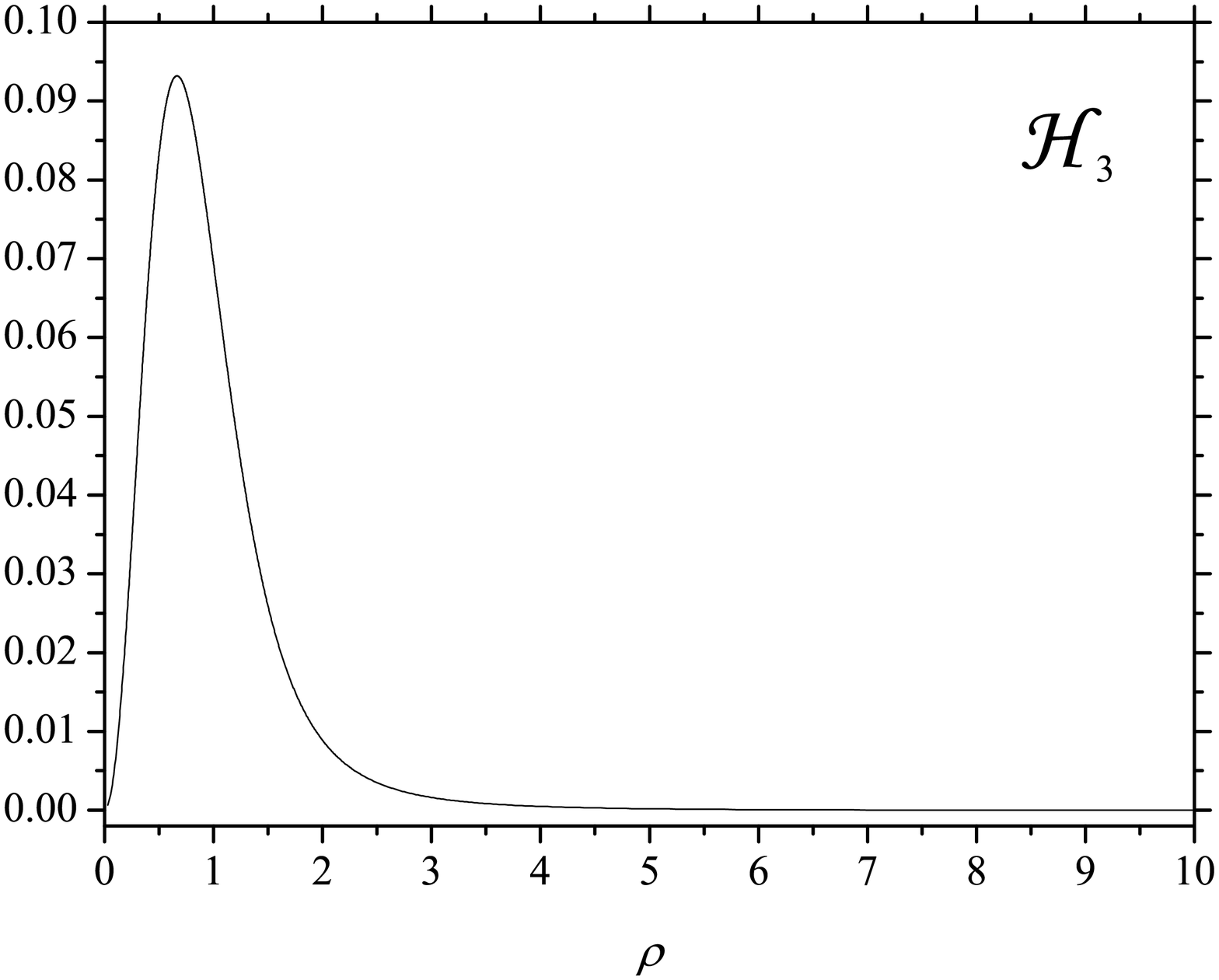}\\
\hspace{-1cm}\includegraphics[width=8cm,clip]{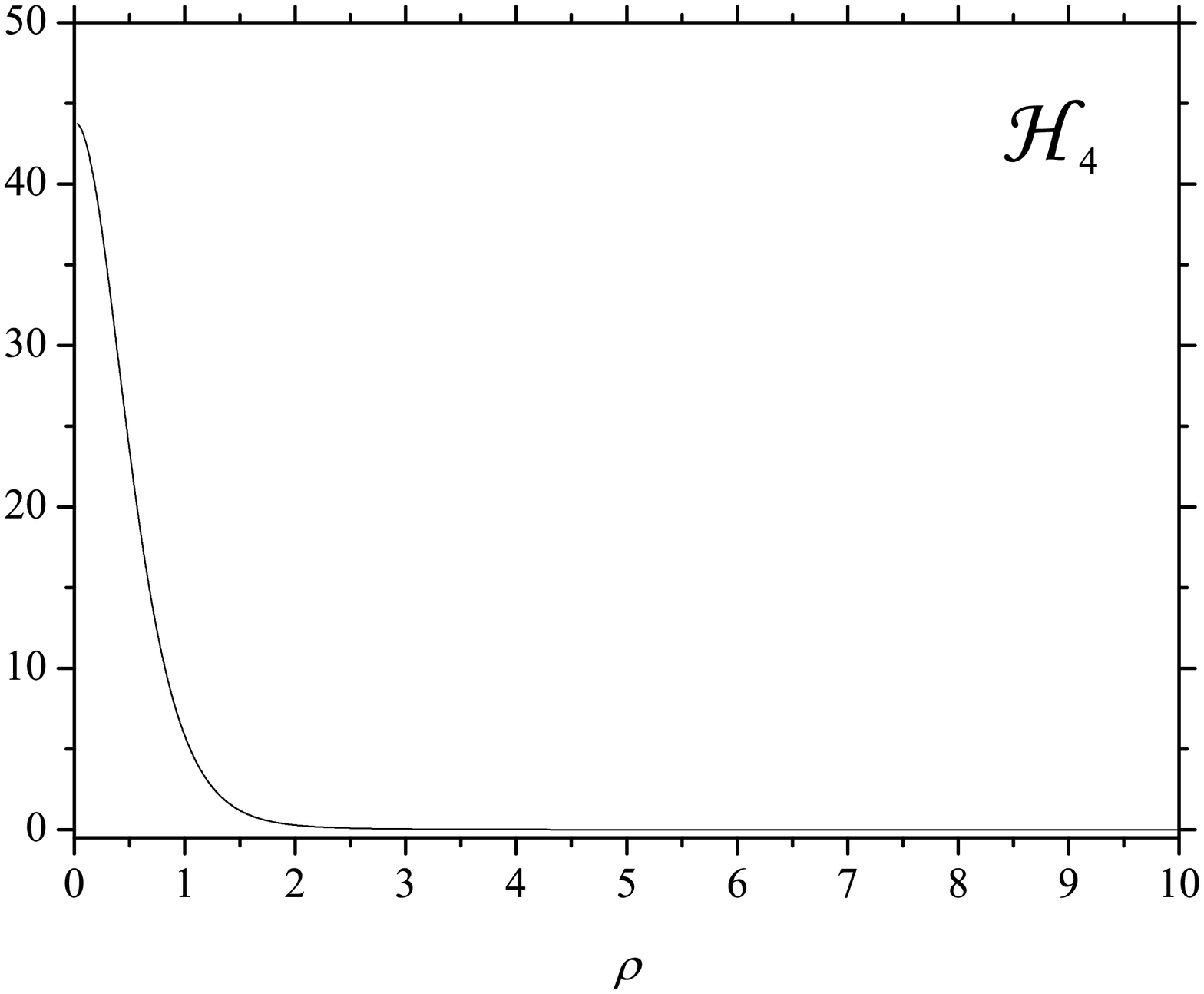}
\includegraphics[width=8cm,clip]{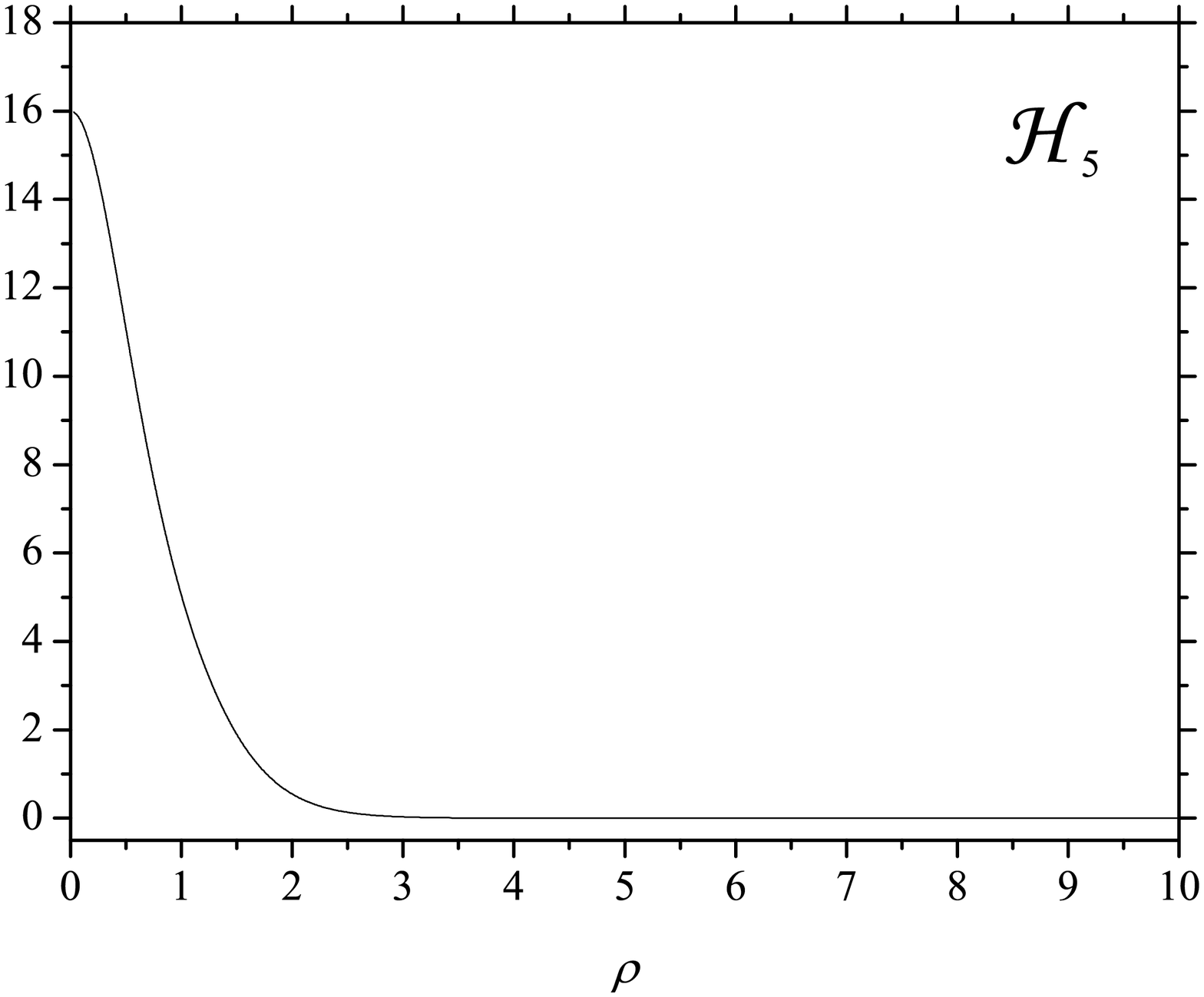}

\caption{\label{energy}The hamiltonian and its components corresponding to the 
solutions of Fig.\ref{profile1} with $k_1=k_2=0.1$.
}
\end{figure}

\subsection{The energy}
One can check that under assumptions about the form of the solution the Lagrangian reduces to the expression that depends only on the radial functions
\br
\mathcal{L}&=&\frac{M^2}{r_0^2}\left(\theta+\frac{\omega}{\rho^2}\right)+\frac{2}{r_0^4e^2}(\beta e^2+\gamma e^2-2)\left(\theta+\frac{\omega}{\rho^2}\right)^2\\
&+&\frac{4}{r_0^4e^2}(\gamma e^2-1)\frac{\zeta^2-\theta\omega}{\rho^2}+\frac{2}{r_0^4e^2}(\gamma e^2+2)\left(\theta^2+\frac{\omega^2}{\rho^4}-2\frac{\zeta^2}{\rho^2}\right)-\mu^2V\,.\nonumber
\er
The first term is just a $CP^N$ Lagrangian. The last two terms, which are proportional to $\gamma e^2-1$ and $\gamma e^2+2$, vanish for the holomorphic solutions since the constraint $\partial_{\mu}u_i\partial^{\mu}u_j=0$ leads to the relation $f'_j(\rho)=\frac{n_j}{\rho} f_j(\rho)$ resulting in equalities $\theta=\frac{\omega}{\rho^2}=\frac{\zeta}{\rho}$ and $\xi=\frac{\eta}{\rho}$. The Hamiltonian is defined by the formula
\br
\mathcal{H}&=&\frac{\delta\, {\cal L}}{\delta\,\partial_{0}u_i}\,\partial_0 u_i 
+\frac{\delta\, {\cal L}}{\delta\,\partial_{0}u_i^*}\,\partial_0 u_i^* 
- {\cal L}\nonumber\\
&:=&\mathcal{H}_1+\mathcal{H}_2+\mathcal{H}_3+\mathcal{H}_4+\mathcal{H}_5
\lab{hamiltonian}
\er
where the components (in the unit of $-M^4e^2/4$) are given by
\br
\mathcal{H}_1=-\left(\theta+\frac{\omega}{\rho^2}\right)\,,~~~~
\mathcal{H}_2=-2\left(\frac{d\psi}{dy}\right)^2\chi\,,~~~~
\mathcal{H}_5=\tilde{\mu}^2 V
\er
\br
\mathcal{H}_3=2\left(\frac{d\psi}{dy}\right)^2\left\{(\beta e^2-1)\left(\theta+\frac{\omega}{\rho^2}\right)\chi+(\gamma e^2-1)\frac{2\eta^2}{\rho^2}\right\}
\er
\br
\mathcal{H}_4&=&\frac{1}{2}(\beta e^2+\gamma e^2-2)\left(\theta+\frac{\omega}{\rho^2}\right)^2+(\gamma e^2-1)\frac{\zeta^2-\theta\omega}{\rho^2}\nonumber\\
&+&\frac{1}{2}(\gamma e^2+2)\left(\theta^2+\frac{\omega^2}{\rho^4}-2\frac{\zeta^2}{\rho^2}\right)-6\left(\frac{d\psi}{dy}\right)^2\left(\xi^2-\frac{\eta^2}{\rho^2}\right)
\lab{hamiltonian_def}
\er
where $\tilde{\mu}^2=r_0^2\mu^2/M^2$.
The reason for splitting the Hamiltonian is partially historical. In general, 
it is a good idea to isolate the contributions that were present 
in the earlier study of the holomorphic vortex solutions i.e $\mathcal{H}_1$, 
$\mathcal{H}_2$ and $\mathcal{H}_3$. The term which has not been present before 
is $\mathcal{H}_4$. Such a term was absent due to the constraint and the 
restriction on the coupling constants $\beta e^2+\gamma e^2=2$.

\section{The numerical analysis}
For the numerical study, it is more convenient to use a new radial coordinate $x$,  defined by
$\rho=\sqrt{\frac{1-x}{x}}$. Accordingly we adopt profile functions $g_i$, instead of using $f_i$, i.e.,
$f_i(\rho)=\sqrt{\frac{1-g_i(x)}{g_i(x)}}$. 
The computations are performed using a standard technique for differential equations, 
the successive over relaxation method.

In the absence of the potential i.e. $\mu^2=0$, the holomorphic solutions \cite{fk} are also solutions of 
the present system. The profile functions of the holomorphic solutions can be written in the coordinate $x$ as
\begin{eqnarray}
g_i(x)=\frac{(\alpha_i^2 x)^{n_i}}{(\alpha_i^2 x)^{n_i}+(1-x)^{n_i}},~~i=1,2
\lab{zerocurvature}
\end{eqnarray}
where $\alpha_i$ are some arbitrary parameters. Our starting point is to reproduce numerically the holomorphic exact solutions 
the \rf{zerocurvature}.
Fig.\ref{profile0} shows the numerical profile functions for $(n_1,n_2)=(1,1)$ and $(1,2)$.

Considering the solutions which do not belong to the integrable sector, we need to set several parameters of the model and the potential as well. 
We employ the potential \rf{potential} with $(a,b)=(0,4)$. This type of the potential has already been 
used for the extended $CP^1$ Skyrme-Faddeev model and it is a potential for the charge $n=1$ integrable sector.  
For the case of 
$(n_1,n_2)=(1,2)$ and $\beta e^2=\gamma e^2=2.0, \tilde{\mu}^2=1.0$, the result is plotted in Fig.\ref{profile1}.
For the energy, we need to determine the functional form $\psi(y)$ and here we simply put as $\psi(y)\equiv y$.
Fig.\ref{energy} is the corresponding hamiltonian density $\mathcal{H}$ and its components 
\rf{hamiltonian}-\rf{hamiltonian_def}.

The following topics should be accented:
\begin{itemize}
\item[(i)]~In this Letter, we have discussed only the cases $(n_1,n_2)=(1,1)$ and $(1,2)$. 
Of course it is expected to exist the solutions with several higher charges. 
\item[(ii)]~For the potential, our definition of the $\mathcal{N}$ factor in~\rf{nfac} is not unique. 
It might be better to write in terms of the principal variable $X(\rho)$. It is mandatory to discuss the several 
possibilities of definition of the potential in the further study. Also it is useful to specify the order
of the potentials $(a,b)$ admitting the existence of the solution. 
\item[(iii)]~The simplest form $\psi(y)=y$ has been applied. In the case of restriction of the solution to the dependence of 
exactly one light-cone coordinate there is a freedom of choice of function $\psi(y)$. Of course, not all possible functions $\psi$ 
lead to acceptable energy densities. In the case of both light cone coordinates $(y_+,y_-)$ the function $\psi$ should be some linear combination of them. 
Unfortunately, in such a case equations of motion become much more complicated. This case shall be studied later.
\item[(iv)]~The $CP^1$ model has the holomorphic solutions with the constraint imposed on the parameters of the model ~\cite{vortexlaf}, 
and also has such solutions without the constraint by introducing a special potential~\cite{fjst}. 
For the $CP^2$ target space, there are the holomorphic solutions in the integrable sector~\cite{fk}. 
Thus it seems natural that there are the holomorphic solutions in the present model with an unique form of the potential. 
\end{itemize}
We will report on these issues in more detailed article.
\vspace{0.5cm}

{\bf Acknowledgement}
The authors would like to thank L. A. Ferreira for discussions and comments.
We are grateful to Kouichi Toda for useful discussions. 
This work was financially supported by a grant of Heiwa Nakajima Foundation 
especially for stay of Pawe\l~Klimas in Japan.

\end{document}